# Deterministic switching of perpendicularly magnetic layers by spin orbital torque through stray field engineering

Sumei Wang[1], Meiyin Yang[2], and Chao Zhao[1]

1. ICAC, Institute of Microelectronics of CAS, University of CAS, No.3 BeiTuCheng West Rd., Beijing 100029, China
2. SKLSM, Institute of Semiconductors of CAS, P. O. Box 912, Beijing 100083, China

We proposed a novel multilayer structure to realize the deterministic switching of perpendicularly magnetized layers by spin orbital torque from the spin Hall effect through stray field engineering. In our design, a pinned magnetic layer is introduced under the heave metal separated by an insulator, generating an in-plane stray field in the perpendicularly magnetized layer. We have confirmed the deterministic switching of perpendicularly magnetized layers through micromagnetic simulation and theoretical analysis. The in-plane stray field accounts for the deterministic switching exhibited in the structure and the reversal ultimate state of the magnetic layer is predictable when the applied spin current density is above the critical spin current density. Moreover, the stray field is easily tunable in a wide range by adjusting the saturation magnetization and dimensions of the pinned layer, and can accommodate different perpendicularly magnetized materials without any external magnetic field.

Extensive experiments have been devoted to study the deterministic switching of perpendicularly magnetized layers in heavy metal/ferromagnet devices [1-10]. Particularly, the spin orbital torque (SOT) induced by the spin Hall effect (SHE) is one of the most promising candidates for next-generation memory devices due to many advantages such as low power consumption and absence of current leakage. The SOT may induce the rotation of magnetization in the magnetic layer, but the final state of the magnetic layer (pointing up or down) is uncertain [11, 12]. Several attempts have been made to eliminate this uncertainty. In Liu's experiment[12], a fixed external field was applied along the charge current direction. The external field succeeded in breaking the symmetry of the rotation in response to the SOT and deterministic switching is achieved. Other alternatives are also put forward to circumvent the complexity of applying an external field, such as introducing antiferromagnetic interaction[10], establishing a lateral structure asymmetry[6], constructing a hybrid ferromagnetic/ferroelectric structure[7], etc. However, the range of assisted field is relatively limited or the design is not easily scalable[8]. Here, we proposed a novel structure to accomplish the deterministic switching of the perpendicular magnetized layer. In the design, a fixed or pinned layer is introduced at the bottom of the magnetic layer and its stray field on the magnetic layer servers as assisted field to break the symmetry. We have confirmed the deterministic switching behavior and feasibility of our design through micromagnetic simulation and theoretical analysis, and the detailed dynamics during switching are also analyzed. The stray field the pinned layer produces on the magnetic layer is independent of the exchange interaction with the magnetic layer, which facilitates the design and application of such a structure. The magnitude of the stray field can be tuned by altering the magnetization and dimensions of the pinned layer. Besides, pinned magnetic layers have been extensively used in industries and are anticipated to be easily implemented.

Illustrations of our design of the multilayer structure can be seen in Fig.1 (a). The key lies in the introduction of the pinned magnetic layer at the bottom of the spintronic device. The magnetization of the fixed layer or the pinned layer is along -*y* direction (in-plane) and hence produces a stray field in the opposite direction (+*y*) on the magnetic layer. To be simple, we refer to the stray field produced by the pinned layer on the magnetic layer as the stray field. The pinned layer is separated by an insulator layer from the magnetic layer. The SOT is a short-range interaction since the torque is caused by spin-orbit interaction, and hence the magnetization of the pinned layer is hardly affected by the SOT due to the separation of the insulator layer. The dynamics of the reversal process are difficult to capture in experiments, but readily accessible by micromagnetic simulation. The dynamics of magnetization are described by the LLG equation with the SOT induced by the SHE, shown in Eq. (1).

$$\frac{d\hat{m}}{dt} = -\gamma \hat{m} \times \vec{H}_{\text{eff}} + \alpha \hat{m} \times \frac{d\hat{m}}{dt} + \frac{\hbar J_s}{2etM_s}\hat{m} \times \hat{\sigma} \times \hat{m} \qquad (1)$$

The last term can be combined with the effective field $H_{\text{eff}}$ equivalently after a simple derivation and we define this field-like SOT term as $\vec{H}_{\text{SOT}} = \frac{\hbar J_s}{2eM_s t}\hat{m} \times \hat{\sigma}$. This field-like term is always perpendicular to the magnetization. Therefore, the total field becomes $\vec{H}'_{\text{eff}} = \vec{H}_{\text{eff}} + \vec{H}_{\text{SOT}} = \vec{H}_a + \vec{H}_{\text{st}} + \vec{H}_{\text{SOT}}$ where $\vec{H}_a$ is the effective anisotropy field, and $\vec{H}_{\text{st}}$ is the stray field. Due to small distances on the scale of nanometers, the dipole approximation fails to apply here, and the stray field needs be evaluated by Eq.(2)[13].

$$\vec{H}_{\text{st}} = \iiint d^3\vec{r}' [-\vec{\nabla}' \cdot \vec{M}(\vec{r}')] \frac{\vec{r} - \vec{r}'}{|\vec{r} - \vec{r}'|^3} \qquad (2)$$

The stray field can be analytically acquired by Eq.(2) since the magnetization of the pinned layer is assumed to be fixed, which can be basically guaranteed by high in-plane anisotropy or strong antiferromagnetic coupling to weaken the influence of the stray field by the magnetic layer. Typical magnetic parameters are chosen for the perpendicular magnetic layer in our calculation: the saturation magnetization $M_s$ is 800 emu/cm$^3$, the effective anisotropy constant $K_{\text{eff}}$ is 1×10$^6$ erg/cm$^3$, and the exchange stiffness is 1×10$^{-6}$ erg/cm within the magnetic layer. The saturation magnetization of the pinned layer $M_{\text{sp}}$ is chosen as 400 emu/cm$^3$ unless otherwise stated. Our simulation is constructed as follows: a charge current is applied along *y*-direction and generates a spin polarization pointing in *x*-direction ($\hat{\sigma}$); the magnetic layer has an out-of-plane easy axis (*z*). The stray field is calculated by Eq.(2) and applies on the magnetic layer. Under the stray field from the pinned layer and without any currents, the magnetization $\hat{m}$ is slightly tilted to *y* axis (about 11°), but no preference of pointing up ($m_z$ >0) or down ($m_z$ <0) is shown. It is worthy to mention that since the magnetization of the pinned layer is in-plane, it will be easy to distinguish the magnetization between the magnetic layer and pinned layer during magnetization characterization.

We apply the spin current after the multilayer structure relaxes to an equilibrium state to investigate the switching properties of the magnetic layer. The spin current density is varied from -22MA/cm² to 22MA/cm² or oppositely swept. Fig.2 presents the switching loops with different $M_{sp}$ when the initial magnetization of the magnetic layer is pointing up ($m_z$ >0). The critical spin current density $J_c$ is 18MA/cm² and 20MA/cm² when the saturation magnetization $M_{sp}$ is 400emu/cm³ and 318emu/cm³, respectively, in agreement with Ref.[12]. We obtained exactly the same switching loop when the initial magnetization is pointing down ($m_z$ <0) (not shown here). Therefore, the ultimate state of the magnetic layer is determinant regardless of the prescribed different initial conditions: pointing down (-*z*) is favorable for positive spin currents, while pointing up (+*z*) is preferable for negative spin currents. Clearly, opposite favoring will be obtained if the initial magnetization of the pinned layer is reversed.

It is crucial to study the dependence of the stray field on the dimensions of the pinned layer and distance between the magnetic layer and pinned layer to adjust the distribution and magnitudes of the stray fields. Fig.3 shows the normalized distribution of the stray field when the size of the pinned layer differs. When the area of the magnetic layer $A_m$ is the same with that of the pinned layer $A_p$, striking non-uniformity in the stray field is exhibited, as seen in Fig.3(a). Furthermore, the perpendicular component of the stray field at the corners is dominant when $A_p=A_m$, which probably jeopardizes the thermal stability of the magnetic layer. To rectify this, the pinned layer needs to be expanded, as seen in Fig.3(b). When the area of the pinned layer quadruples while the magnetic layer is kept centered on the pinned layer, the imbalance is substantially improved whilst the *x* and *z* components of the stray field are about one order smaller than the *y* component. Note that the magnitude of stray field declines in Fig.3(b) compared to that in Fig.3(a), but the reduction can be compensated by increasing $M_{sp}$, which is linearly proportional to the stray field. Additionally, the stray field is sensitive to the thickness of the pinned layer. The stray field almost doubles when we raise the thickness of the pinned layer from the 2nm to 4nm. Therefore, the stray field can be adjusted in a wide range and is flexible to accommodate different perpendicularly magnetized materials. Another interesting finding is that the stray field is quite insensitive to the vertical distances between the pinned layer and magnetic layer within nanometer scale. For instance, only a slight difference (2%) in the stray field is found if the distance increases from 4.4nm to 6.4nm.

Periodic spin current pulse of ±0.5 ns is applied to orient the magnetic layer to investigate its switching behavior, and the results are displayed in Fig.4 where the damping constant is 0.5. There magnitudes of the spin current density were studied:

20MA/cm$^2$, 25MA/cm$^2$ and 30MA/cm$^2$. The results indicate that the magnetic layer is successfully switched when reversal spin currents are applied ($J_s \neq 0$). After switching, a relaxation process ensues ($J_s = 0$). When the current is turned off, significant jumps in the variation of average magnetization (dent or bump in Fig.4(b)) are observed but the magnetization rapidly reaches an equilibrium state. Note that the stray field always exists and the ultimate magnetization is still slightly tilted. However, the small deviation of the magnetization from its easy axis will not affect its overall orientation (up or down), i.e., the stored information is unchanged during the relaxation process. Although successful switching is also achieved for 25MA/cm$^2$ and 30MA/cm$^2$, the change of $|m_z|$ is different (significant for 30MA/cm$^2$ case). Before the spin current is off, $|m_z|$ remains steady at different values: $|m_z|$=0.97 for 20MA/cm$^2$, $|m_z|$=0.93 for 25MA/cm$^2$ and $|m_z|$=0.32 for 30MA/cm$^2$. The spin orbital torque is larger at larger spin currents, and leads to the magnetization rotating to the *y* axis. In fact, for 30MA/cm$^2$, the magnetization of the magnetic layer is dominantly prone to lie in *x-y* plane but still relaxes to the expected orientation when the spin current is off.

Evidently, the switching trajectory and behavior are closely related to the damping constant and duration of applied currents and relaxation process, especially with lower damping constants[14]. We have found that oscillation with gradually decreasing amplitudes occurs when the damping reduces and the switching and relaxation may not reach the equilibrium states within prescribed time. For example, the switching process is not strictly repetitive in Fig. 5(a) where the damping constant is 0.1. In Fig.5(b), we have increased the period to 6ns, successful and repetitive dynamic reversal are obtained. We also examined the switching behaviors when the damping constant is 0.02 and 0.05. As expected, unsuccessful switching occurs in some extreme cases. In other words, it should be taken into account in actual applications that relatively slower switching needs to be atoned by longer duration of applied currents and relaxation processes, and further studies are required.

To conclude, we have proposed a novel and simple design to realize the deterministic switching of perpendicularly magnetized layers via manipulating the stray field generated by a pinned layer. We have verified the deterministic switching by micromagnetic simulation and theoretical analysis. The stray field breaks the symmetry of rotation responding to the SOT induced by the SHE and accounts for the deterministic switching of the magnetic layer in the structure. The stray field is easily tunable and flexible to accommodate different perpendicularly magnetized materials without any external magnetic field, on the basis that the stray field is proportional to the saturation magnetization of the pinned layer and sensitive to its dimensions.

The authors would like to thank Prof. L.Q. Liu, Prof. D. Wei, and Dr. Y. Wang for valuable discussion.

## References

[1]  I. M. Miron, K. Garello, G. Gaudin, P. J. Zermatten, M. V. Costache, S. Auffret, *et al.*, "Perpendicular switching of a single ferromagnetic layer induced by in-plane current injection," *Nature,* vol. 476, pp. 189-194, Aug 11 2011.

[2]  L. Q. Liu, C. F. Pai, Y. Li, H. W. Tseng, D. C. Ralph, and R. A. Buhrman, "Spin-Torque Switching with the Giant Spin Hall Effect of Tantalum," *Science,* vol. 336, pp. 555-558, May 4 2012.

[3]  K. Ando, S. Takahashi, K. Harii, K. Sasage, J. Ieda, S. Maekawa, *et al.*, "Electric manipulation of spin relaxation using the spin Hall effect," *Physical Review Letters,* vol. 101, p. 036601, Jul 18 2008.

[4]  T. Suzuki, S. Fukami, N. Ishiwata, M. Yamanouchi, S. Ikeda, N. Kasai, *et al.*, "Current-induced effective field in perpendicularly magnetized Ta/CoFeB/MgO wire," *Applied Physics Letters,* vol. 98, p. 142505, Apr 4 2011.

[5]  L. You, O. Lee, D. Bhowmik, D. Labanowski, J. Hong, J. Bokor, *et al.*, "Switching of perpendicularly polarized nanomagnets with spin orbit torque without an external magnetic field by engineering a tilted anisotropy," *Proceedings of the National Academy of Sciences of the United States of America,* vol. 112, pp. 10310-10315, Aug 18 2015.

[6]  G. Q. Yu, P. Upadhyaya, Y. B. Fan, J. G. Alzate, W. J. Jiang, K. L. Wong, *et al.*, "Switching of perpendicular magnetization by spin-orbit torques in the absence of external magnetic fields," *Nature Nanotechnology,* vol. 9, pp. 548-554, Jul 2014.

[7]  K. Cai, M. Yang, H. Ju, S. Wang, Y. Ji, B. Li, *et al.*, "Electric field control of deterministic current-induced magnetization switching in a hybrid ferromagnetic/ferroelectric structure," *Nature Materials,* 2017.

[8]  Y. C. Lau, D. Betto, K. Rode, J. M. D. Coey, and P. Stamenov, "Spin-orbit torque switching without an external field using interlayer exchange coupling," *Nature Nanotechnology,* vol. 11, pp. 758-762, Sep 2016.

[9]  Y. W. Oh, S. H. C. Baek, Y. M. Kim, H. Y. Lee, K. D. Lee, C. G. Yang, *et al.*, "Field-free switching of perpendicular magnetization through spin-orbit torque in antiferromagnet/ferromagnet/oxide structures," *Nature Nanotechnology,* vol. 11, pp. 878-885, Oct 2016.

[10] S. Fukami, C. L. Zhang, S. DuttaGupta, A. Kurenkov, and H. Ohno, "Magnetization switching by spin-orbit torque in an antiferromagnet-ferromagnet bilayer system," *Nature Materials,* vol. 15, pp. 535-542, May 2016.

[11] M. Yang, K. Cai, H. Ju, K. W. Edmonds, G. Yang, S. Liu, *et al.*, "Spin-orbit torque in Pt/CoNiCo/Pt symmetric devices," *Scientific Reports,* vol. 6, 2016.

[12] L. Q. Liu, O. J. Lee, T. J. Gudmundsen, D. C. Ralph, and R. A. Buhrman, "Current-Induced Switching of Perpendicularly Magnetized Magnetic Layers Using Spin Torque from the Spin Hall Effect," *Physical Review Letters,* vol. 109, p. 096602, Aug 29 2012.

[13] D. Wei, S. Wang, Z. Ding, and K. Z. Gao, "Micromagnetics of ferromagnetic nano-devices using the fast Fourier transform method," *IEEE Transactions on Magnetics,* vol. 45, pp. 3035-3045, 2009.

[14] Y. Huai, "Spin-Transfer Torque MRAM (STT-MRAM): Challenges and Prospects," *AAPPS Bulletin,* pp. 33-40, 2008.

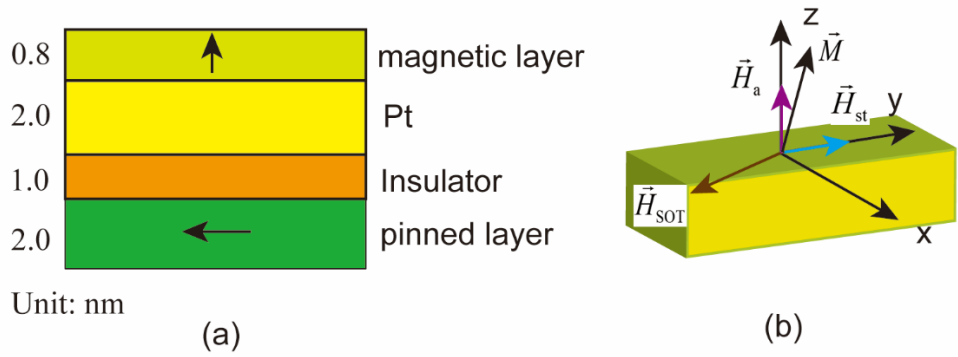

Fig.1 (a) Schematics of our design: the magnetization of pinned layer is assumed to be along *y*-axis (b) Illustrations of the effective field: the stray field is dominantly along *y* axis and the initial magnetization of the magnetic layer is tilted towards *y* axis by a degree of 11° within *y*-*z* plane

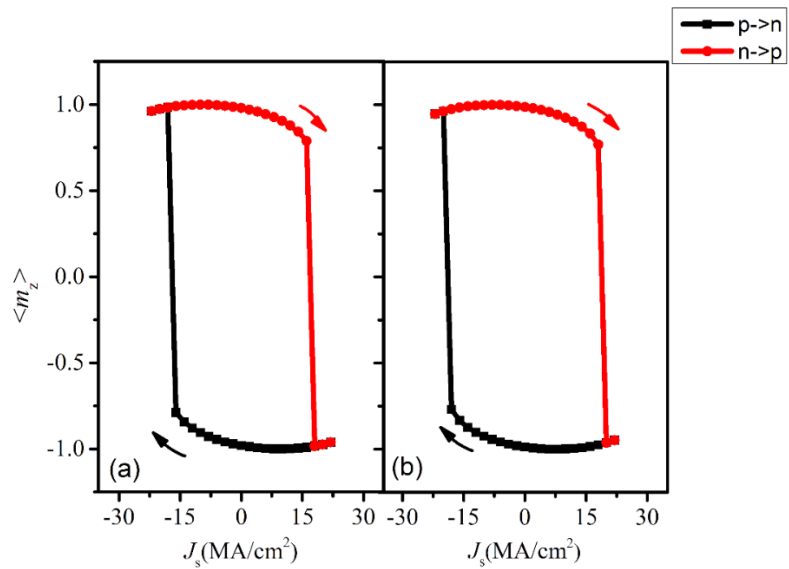

Fig. 2. The switching loops of the magnetic layer: (a) $M_{sp}$=400emu/cm$^3$ (b) $M_{sp}$=318emu/cm$^3$. In the figure, n and p represent negative (-*z*) and positive (+*z*) spin current, respectively, and $m_z$ is the average of normalized *z* component of magnetization in the magnetic layer

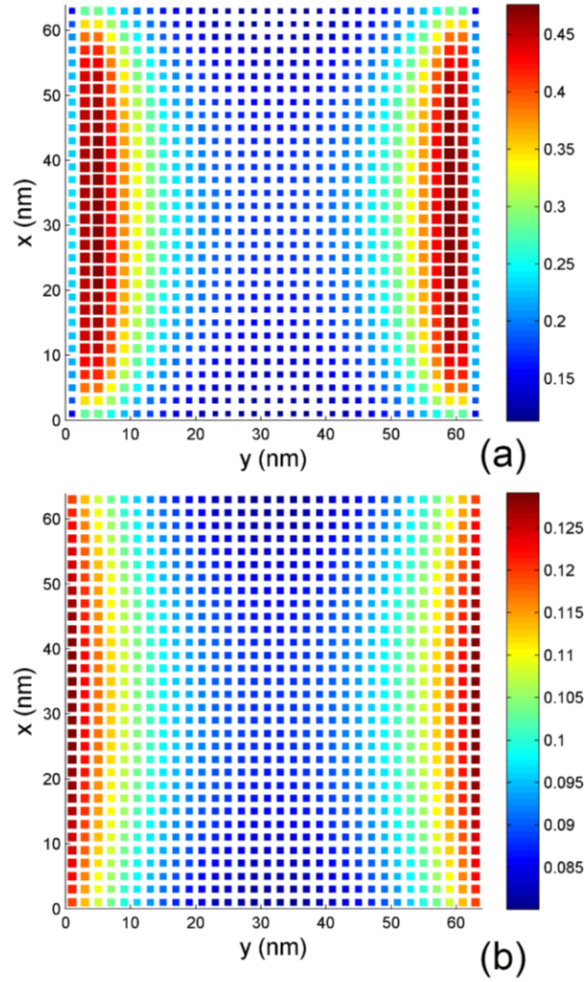

Fig.3 Distribution of the stay field located at the magnet layer normalized by $4\pi M_{sp}$: (a) $A_p=A_m$ (b) $A_p=4A_m$. The square size is proportional to the $y$ component of the stray field.

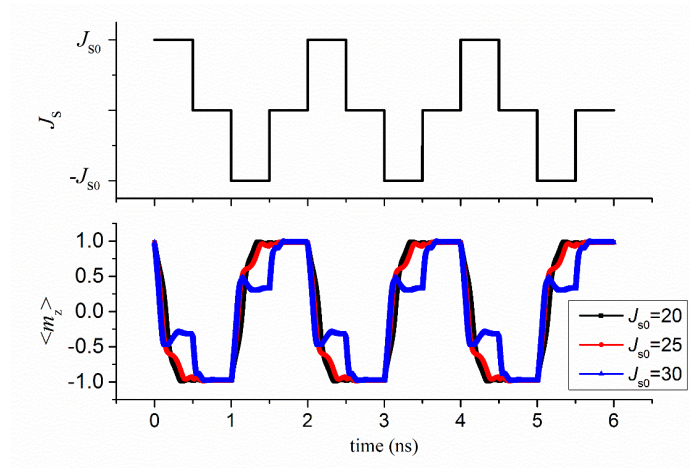

Fig.4 Variation of average magnetization under periodic spin current density: (a) the applied spin current density, (b) the variation of the average magnetization of the magnetic layer

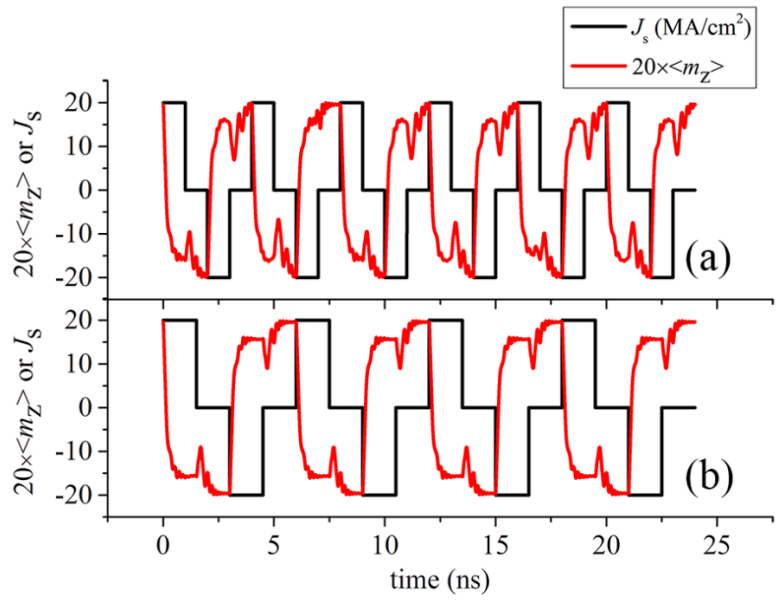

Fig.5 Variation of average magnetization under periodic spin current density with different spin current duration when $J_{s0}$=20MA/cm$^2$: (a) the period is 4ns (b) the period is 6ns. Note that the normalized magnetization is multiplied by $J_{s0}$ for visual effect

8